# Hidden Symmetry and Mutiferroicity in a Triangular Spin Lattice With Proper-Screw-Spin Chain Order

E. J. Kan[1], H. J. Xiang[2], Y. Zhang[1], C. Lee[1] and M.-H. Whangbo[1]*

[1] Department of Chemistry, North Carolina State University, Raleigh, North Carolina 27695-8204

[2] National Renewable Energy Laboratory, Golden, Colorado 80401

## Abstract

The multiferroicity in a triangular spin lattice with proper-screw-spin chain order was explored by performing density functional calculations for $AgCrO_2$ and analyzing the symmetry of the magnetic structure of the triangular spin lattice. Strong geometric spin frustration exists within and between $CrO_2$ layers, and the ferroelectric polarization originates from the spiral-spin chain structures propagating across the proper-screw-spin chains. The triangular spin lattice can have ferroelectric polarization parallel to its mirror plane due to hidden symmetry.

Pacs: 75.80.+q, 77.80.-e, 71.20.-b

Multiferroic and magnetoelectric materials have attracted much attention because of their potential applications in memories and sensors as well as fundamental physics questions they raise [1-5]. In multiferroics driven by magnetic order, ferroelectric (FE) polarization appears only in an ordered magnetic state that removes their inversion symmetry. The magnetic-order-driven FE polarization in collinear-spin systems is discussed in terms of exchange striction [6-8], and that in noncollinear-spin systems in terms of spin-orbit coupling (SOC) [9-11]. Qualitatively, the FE polarization of a spiral-spin chain is described by the Katsura-Nagaosa-Balatsky (KNB) model [12], according to which a chain of cycloid spin order (hereafter, a cycloid-spin chain) gives rise to FE polarization, but a chain of proper-screw spin order (hereafter, a screw-spin chain) (**Fig. 1a**) does not. The magnetic oxide $AgCrO_2$ consists of $CrO_2$ trigonal layers made up of edge-sharing $CrO_6$ octahedra with $Cr^{3+}$ ($d^3$, $S = 3/2$) ions (**Fig. 1b**). Each $CrO_2$ layer can be subdivided into chains of edge-sharing $CrO_6$ octahedra running along the *a*, *b* or (*a+b*) direction. Adjacent $CrO_2$ layers are stacked along the *c* direction with diamagnetic $Ag^+$ ($d^{10}$) ions intercalated between the $CrO_2$ layers to form linear O-Ag-O bridges such that a repeat unit cell has three $CrO_2$ layers (**Fig. 1c**). $AgCrO_2$ exhibits FE polarization in the magnetic ground state below ~21 K, in which each triangular spin lattice (TSL) of $Cr^{3+}$ ions separates into screw-spin chains of $Cr^{3+}$ ions [13,14]. This finding is unexpected from the viewpoint of the KNB model. The FE polarization in $AgCrO_2$ and its isostructural analogue $CuFeO_2$ [15-17] has been explained on the basis of the finding that a magnetic structure with $C_2$ rotation symmetry leads to FE polarization along the $C_2$ rotation axis [18]. Arima examined the possible directions of FE polarization in $CuFeO_2$ by analyzing the symmetry of the magnetic structure of an isolated $FeO_2$ layer with screw-spin chain order [17]. His analysis showed that the direction of the FE polarization depends on the orientation of the screw-rotation plane. Seki *et al.* presented a similar discussion for the FE polarization of $AgCrO_2$ [14].

In understanding the screw-spin chain order of $AgCrO_2$, it is critical to know its spin exchange interactions, which include the intra-layer spin exchange $J_{NN}$, $J_{NNN}$ and $J_{ab}$ and the inter-layer spin exchange $J_c$ (**Fig. 1c**). In general, non-collinear spin arrangements occur to reduce the extent of geometric spin frustration [19]. If the spin exchange other than $J_{NN}$ is negligible, each $CrO_2$ layer should have the noncollinear 120° magnetic structure [19]. The screw-spin chain order that occurs in the $CrO_2$ layers leads to a non-collinear spin arrangement within and between adjacent $CuO_2$ layers, thereby implying the existence of geometric spin frustration within and between adjacent $CrO_2$ layers. It is of interest to verify this implication by evaluating the values of $J_{NN}$, $J_{NNN}$, $J_{ab}$ and $J_c$ on the basis of electronic structure calculations. Concerning the FE polarization of $AgCrO_2$, its direction was predicted to be along a $C_2$ axis by symmetry analysis [14,17]. It is of interest to see if $AgCrO_2$ can have FE polarization along other symmetry-dictated directions. In addition, electronic structure calculations are necessary to describe both the direction and the magnitude of the FE polarization. In this Letter, we investigate these questions by performing first-principles density functional theory (DFT) calculations for $AgCrO_2$ with and without including spin-orbit coupling (SOC) effects and also by analyzing the symmetry of the magnetic structure of its TSL.

Our spin-polarized DFT calculations were performed by using the projector augmented wave method [20] coded in the Vienna *ab initio* simulation package [21] with the local density approximation (LDA) and the plane wave cut-off energy of 400 eV. The

LDA plus on-site repulsion U method (LDA + U) was employed to properly describe the strong electron correlation associated with the Cr 3d states [22]. The values of U = 2.3 eV and J = 0.96 eV on Cr reported in the literature [23] were adopted for our LDA+U calculations. The electronic structure of $AgCrO_2$ calculated for the ferromagnetic (FM) state is presented in **Fig. 2**, which shows that the FM state is insulating with an indirect band gap. The band dispersion is strong in the *ab* plane but weak along the *c* direction (**Fig. 2a**), as expected for the layered structure of $AgCrO_2$. The plots of the density of states (DOS) and the partial DOS reveal the presence of three electrons in the up-spin Cr 3d bands, which is consistent with the picture of high-spin $Cr^{3+}$ ($d^3$) ions in $AgCrO_2$.

To evaluate the values of $J_{NN}$, $J_{NNN}$, $J_{ab}$ and $J_c$, we determine the total electronic energies of five ordered magnetic states of $AgCrO_2$ [24] by performing LDA+U calculations and express their total spin exchange energies in terms of the spin Hamiltonian $\hat{H} = \sum_{i<j} J_{ij} \hat{S}_i \cdot \hat{S}_j$ written in terms of the four exchange parameters, i.e., $J_{ij}$ = $J_{NN}$, $J_{NNN}$, $J_{ab}$, or $J_c$. Then, by mapping the energy differences between the five ordered magnetic states determined from the LDA+U calculations onto the corresponding energy differences determined from the spin Hamiltonian, we obtain the values of $J_{NN}$, $J_{NNN}$, $J_{ab}$ and $J_c$ presented in **Fig. 1c** [24]. The antiferromagnetic spin exchange decreases in the order, $J_c > J_{NN} \geq J_{NNN}$. Both $J_{NNN}$ and $J_c$ are super-super exchange interactions involving the $t_{2g}$ states of the $CrO_6$ octahedron [25]. $J_{NNN}$ occurs through two Cr-O…O-Cr linkages, and $J_c$ through two Cr-O…Ag…O-Cr linkages. Within each $CrO_2$ layer, spin frustration occurs in every triangle of $J_{NN}$, in every triangle of $J_{NNN}$, and along every edge-sharing chain due to $J_{NN}$ and $J_{NNN}$. Between adjacent $CrO_2$ layers, spin frustration occurs in every isosceles triangle made up of $J_{NN}$ and $J_c$. It is most likely that the screw-spin chain order in the CrO$_2$ layers of AgCrO$_2$ results to minimize the extent of the strong inter- and intra-layer geometric spin frustration.

Experimentally, the propagation vector of the screw-spin chains is found to be (q, q, 0) with q = 1/3 [13,14], namely, the screw-spin chains run along the (*a*+*b*) direction with screw-rotation angle of 120°. Since the (*a*+*b*) direction is equivalent to the *a* or *b* direction in each CrO$_2$ layer (**Fig. 1b**), it will be assumed in our study that the screw-spin chains run along the *a* direction with the screw-rotation angle 120° and clockwise rotation of the spins along the *a* direction. In each CrO$_2$ layer, we generate three ordered magnetic structures described by the propagation vectors $Q_a$ = (1/3, 0, 0), $Q_{a+b}$ = (1/3, 1/3, 0) and $Q_{-a+b}$ = (−1/3, 1/3, 0) (**Fig. 3**). Here the repeat patterns are generated by using the direct vectors (*a*, *b* and *c*) instead of the reciprocal vectors (*a**, *b** and *c**). Thus, in the $Q_a$ state, the screw-spin chains repeat along the *b* direction without changing the screw-rotation angle (**Fig. 3a**). In the $Q_{a+b}$ state, the screw-spin chains repeat along the *b* direction while advancing the screw-rotation clockwise by 120° (**Fig. 3b**). In the $Q_{-a+b}$ state, the screw-spin chains repeat along the *b* direction while advancing the screw-rotation anticlockwise by 120° (**Fig. 3c**).

On the basis of LDA+U calculations, we evaluate the FE polarizations for the $Q_a$, $Q_{a+b}$ and $Q_{-a+b}$ states of AgCrO$_2$ by using the Berry phase method [26] and the (3a, 3b, c) supercell and also by fully relaxing the atom positions. These calculations do not lead to any FE polarization, so the observed FE polarizations of AgCrO$_2$ cannot be explained by the exchange striction mechanism. We then evaluate the FE polarizations of the three states on the basis of LDA+U+SOC calculations using the structures optimized by the

LDA+U calculations. As summarized in **Table 1**, these LDA+U+SOC calculations show substantial FE polarization for the $Q_a$, $Q_{a+b}$ and $Q_{-a+b}$ states. Therefore, as found for the FE polarization in spiral-spin chain systems [9-11], SOC is essential for the FE polarization of the screw-spin states, because it allows the ordered magnetic state with no inversion symmetry to adopt its optimal electron density distribution.

In the $Q_{a+b}$ state, the FE polarization occurs only along the *a* direction (i.e., $P_{//a}$ = 140.6 µC/m$^2$). Given the equivalence of the (*a* + *b*) and *a* directions of the TSL (**Fig. 1b**), this result confirms Arima's symmetry analysis for the case of the (q, q, 0) propagation of the screw-spin chain [17], for which the screw-spin chains are along the (*a*+*b*) direction and the magnetic structure of the TSL has $C_2'$ symmetry (i.e., $C_2$ rotation followed by time reversal) along the chain direction (**Fig. 3b**, **3d**). The calculated value of $P_{//a}$ = 140.6 µC/m$^2$ is greater than the experimental value [14] by a factor of 7. In the $Q_a$ state, the electric polarization is nonzero along the $\perp(a+b)$ direction in the *ab* plane (i.e., $P_{\perp(a+b)}$ = 75.5 µC/m$^2$) (**Fig. 3a**) and also along the *c* direction (i.e., $P_{//c}$ = 32.3 µC/m$^2$). Note that the $\perp(a+b)$ direction is parallel to a mirror plane (**Fig. 1b**). The FE polarization of the $Q_{-a+b}$ state is similar to that of the $Q_a$ state, except that its *ab*-plane component of the FE polarization is perpendicular to the *b* direction (**Fig. 3c**). The $\perp b$ direction in the *ab*-plane is also parallel to a mirror plane of the TSL (**Fig. 1b**). The magnetic structure of the TSL has $C_2'$ symmetry in the $Q_{a+b}$ state (**Fig. 3b**, **3d**), but has no global symmetry in the $Q_a$ or $Q_{-a+b}$ state. This apparently puzzling observation is caused by hidden symmetry, as discussed below.

To explain the substantial FE polarization calculated for the three screw-spin states, it is necessary to consider interactions between the screw-spin chains. In the $Q_{a+b}$ state, the interchain spiral-spin propagates along the $b$ and the $(a+b)$ directions with opposite senses of rotation (**Fig. 3b**). In the $Q_a$ and $Q_{-a+b}$ states, the interchain spiral-spin propagates only in one direction, i.e., the $(a+b)$ direction in the $Q_a$ state (**Fig. 3a**) and the $b$ direction in the $Q_{-a+b}$ state (**Fig. 3c**). To discuss the origin of the FE polarization in the $Q_a$ state, we present in **Fig. 4a** two "ribbons" of edge-sharing $CrO_6$ octahedra along the interchain spiral-spin direction [i.e., the $//(a+b)$ direction], which were taken from the magnetic structure of the TSL in the $Q_a$ state (**Fig. 3a**). For simplicity, such ribbons will be referred to as the $//(a+b)$-ribbons. The NN interactions between the upper and lower $//(a+b)$-ribbons occur along the $a$ and $b$ directions. These inter-ribbon interactions involve the proper-screw spins along the $a$ direction, and the collinear spins along the $b$ direction. According to the KNB rule, $\vec{P} \propto \vec{e}_{ij} \times (\vec{S}_i \times \vec{S}_j)$, these interactions do not generate FE polarization. Consequently, every $//(a+b)$-ribbon, perpendicular to the mirror plane, contributes independently to the FE polarization of the TSL, and all the $//(a+b)$-ribbons are identical in nature. Nevertheless, the spiral-spin pattern in each $//(a+b)$-ribbon presents no apparent symmetry that one might use to predict the direction of FE polarization. Thus, as presented in **Fig. 4b** and **4c**, we decompose the spin vectors **S** at the spin sites of the $//(a+b)$-ribbon into two parts, i.e., the perpendicular component $\mathbf{S}_\perp = \mathbf{S}_x + \mathbf{S}_z$ and the parallel component $\mathbf{S}_{//} = \mathbf{S}_y$, with respect to the mirror plane. Here the coordinate axes are chosen such that the xz-plane contains the $//(a+b)$-ribbon plane with the z axis perpendicular to the ribbon. Then, the $\mathbf{S}_\perp$ components of the spin sites form a cycloid-like spiral-spin chain along the $(a+b)$ direction (**Fig. 4b**), which has the $m'$ symmetry (i.e., mirror-reflection followed by time reversal). The cycloid-like spin chain gives rise to FE polarization along the $c$ direction ($P_{//c} \neq 0$) according to the KNB rule. The $S_{//}$ components of the spin sites form a collinear spin pattern as depicted in **Fig. 4c**, which has the mirror-plane symmetry $m$. The latter requires that the associated FE polarization lie in the mirror plane. This accounts for why the FE polarization of the $Q_a$ state is along the $\perp(a+b)$ direction in the $ab$-plane ($P_{\perp(a+b)} \neq 0$). Note that the mirror plane of the $//(a+b)$-ribbon chain, providing the $m'$ and $m$ symmetry for the $S_\perp$ and $S_{//}$ spin components, shifts its position along the $(a+b)$ direction (**Fig. 4a**) on going from one ribbon chain to its adjacent chain. As a result of this hidden symmetry, the TSL exhibits FE polarization although its magnetic structure has no global symmetry. The FE polarization of the $Q_{-a+b}$ state is similarly explained.

In summary, AgCrO$_2$ possesses strong intra- and inter-layer geometric spin frustration, the FE polarization obtained by LDA+U+SOC calculations agrees reasonably well with experiment, and the FE polarization of a TSL with screw-spin chain order originates from the spiral-spin structures propagating across the screw-spin chains. The FE polarization can occur parallel to the mirror plane of the TSL, although the associated magnetic structure has no global symmetry, because of the hidden symmetry associated with the interchain spiral-spin structure.

**Acknowledgements**

The research was supported by the Office of Basic Energy Sciences, Division of Materials Sciences, U.S. Department of Energy, under Grant No. DE-FG02- 86ER45259, and by the National Energy Research Scientific Computing Center under Contract No. DE-AC02-05CH11231. HJX thanks S. Seki for stimulating discussion.

* The corresponding author (mike_whangbo@ncsu.edu)

Table 1. Ferroelectric polarizations (in units of $\mu C/m^2$) determined for the proper-screw-spin states $Q_a$, $Q_{a+b}$ and $Q_{-a+b}$ of AgCrO$_2$ from LDA+U+SOC calculations

|  | $Q_a$ | $Q_{a+b}$ | $Q_{-a+b}$ |
|---|---|---|---|
| (001) | $P_{\perp(a+b)} = -75.5$ | $P_{//a} = 140.6$ | $P_{\perp b} = 75.5$ |
| (001) | $P_{//(a+b)} = 0.0$ | $P_{\perp a} = 0.0$ | $P_{//b} = 0.0$ |
| [001] | $P_{//c} = 32.3$ | $P_{//c} = 0.0$ | $P_{//c} = 32.3$ |

**Figure captions**

Figure 1.    (Color online) (a) Schematic diagrams of spiral-spin chains with cycloid-spin (top) and proper-screw-spin (down) order. (b) A projection view of an isolated $CrO_2$ layer of $AgCrO_2$ showing the $C_2$ and $S_2$ symmetry axes. (c) A projection view of an isolated $CrO_2$ layer of $AgCrO_2$ showing the intralayer spin exchange paths $J_{NN}$, $J_{NNN}$ and $J_{ab}$ (left), and a perspective view of three consecutive layers of $Cr^{3+}$ ions showing the interlayer spin exchange $J_c$ (right).

Figure 2.    (Color online). Electronic structure calculated for the FM state of $AgCrO_2$: (a) Dispersion relations of the up-spin and down-spin bands (red and blue lines, respectively). (b) Total DOS plot and partial DOS plots for the Ag 4d and Cr 3d states.

Figure 3.    (Color online). Schematic representations of the three magnetic states generated by repeating the screw-spin chains running along the *a* direction: (a) $Q_a$, (b) $Q_{a+b}$ and (c) $Q_{-a+b}$. The number 0, 1 and -1 in the $CrO_6$ octahedra show that the screw-rotation angle in their screw-rotation planes are 0°, 120° and -120°, respectively. (d) Perspective view of the screw-spin chains in the $Q_{a+b}$ state emphasizing the presence of $C_2'$ symmetry.

Figure 4. (Color online). (a) Two ribbon chains of edge-sharing $CrO_6$ octahedra along the $//(a+b)$ direction that contain the interchain spiral-spin structure, which were taken from the magnetic structure of the $Q_a$ state presented in Fig. 3a. (b) The perpendicular spin component, $S_\perp$, at each spin site of an isolated $//(a+b)$ ribbon in the state $Q_a$. (c) The parallel spin component, $S_{//}$, at each spin site of the $//(a+b)$ ribbon in the state $Q_a$.

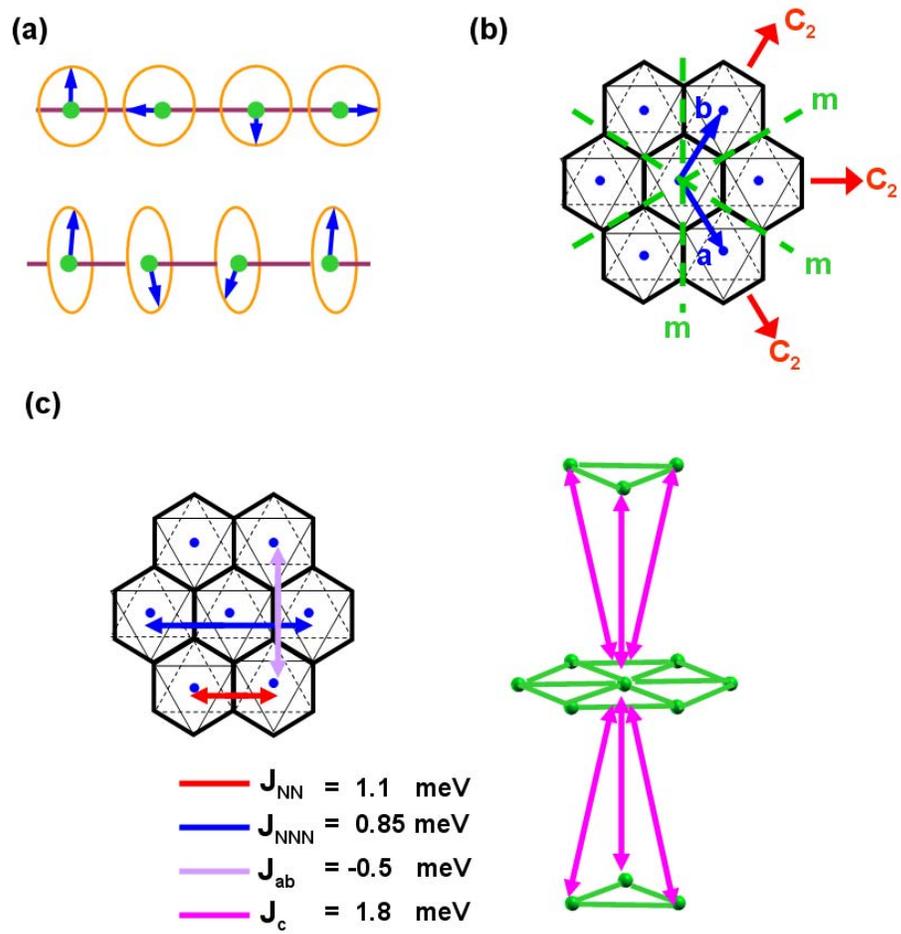

Figure 1

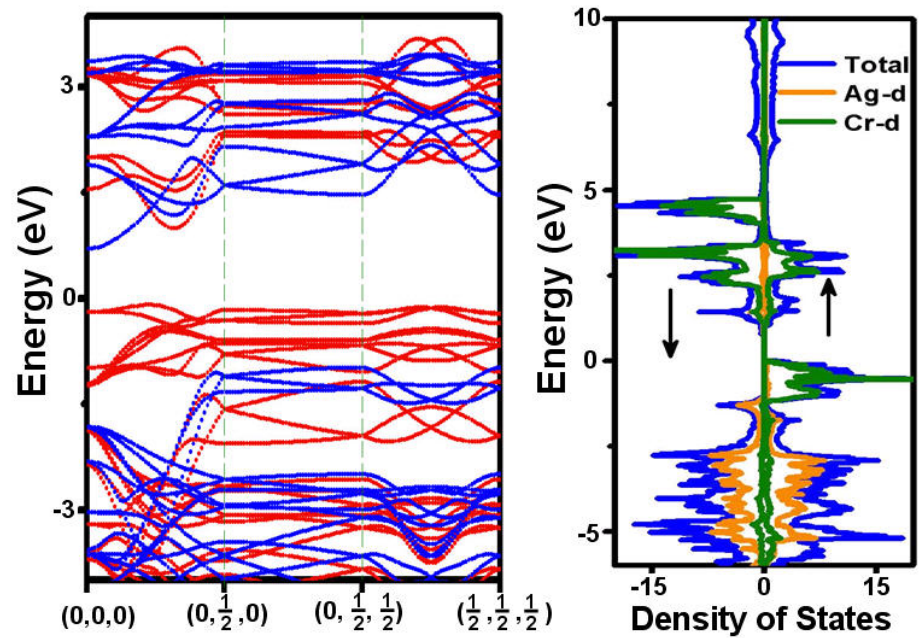

Figure 2

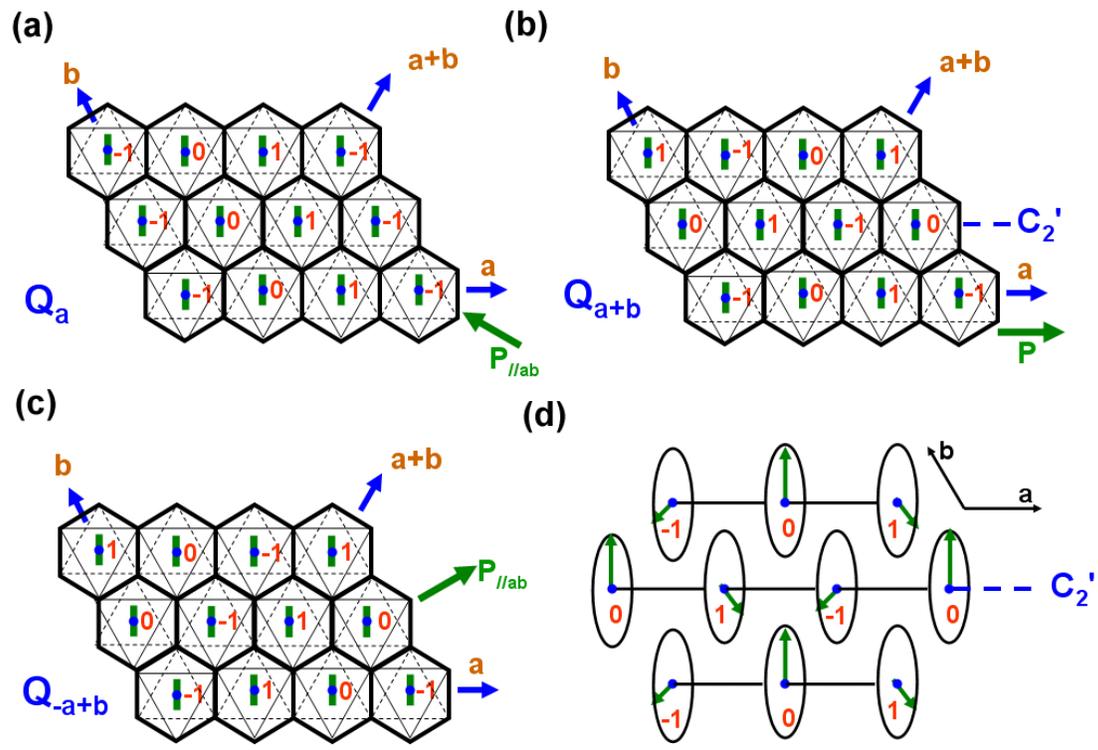

Figure 3

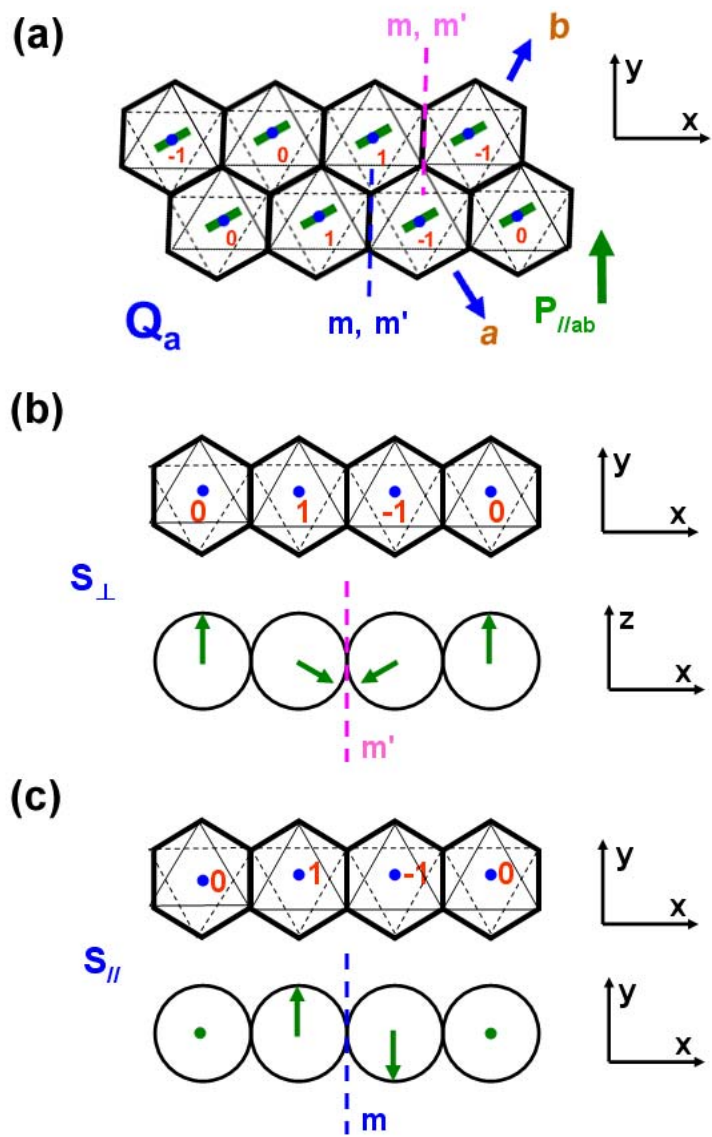

Figure 4.